# STUDY ON PLANAR WHISPERING GALLERY DIELECTRIC RESONATORS. I. GENERAL PROPERTIES

Giuseppe Annino[@], Mario Cassettari, and Massimo Martinelli

Istituto di Fisica Atomica e Molecolare[*], CNR, Area della Ricerca, via Moruzzi 1, 56124 Pisa, Italy.

[*] now Istituto per i Processi Chimico-Fisici

[@] Corresponding author. E-mail: geannino@ifam.pi.cnr.it

## ABSTRACT

Several basic properties of homogeneous planar Whispering Gallery dielectric resonators are investigated through a general approach. In particular the values for the minimum and maximum allowed radius, defined by irradiation and dielectric losses, respectively, are obtained in terms of the complex dielectric constant of the involved media. The same approach is employed in the analysis of the limit allowed frequencies of a fixed family of mode, leading to the definition of the “ideal” frequency band. The role of the transverse mode is then discussed, and the “effective” frequency band is introduced and determined both in simple disc and circular ring resonators. The extension of the effective band is limited at least by a factor 3, in terms of decades, in comparison to the extension of the ideal one; this limitation, due to the presence of transverse modes, can be overcome using suitably designed non-homogeneous resonators, as discussed in a following companion paper.

## 1. Introduction

Planar Dielectric Resonators working on Whispering Gallery Modes (WGDR), starting from their recent discover **[1]**, have assumed a role of increasing importance for applications in the frequency region ranging from microwaves to visible radiation **[2-5]**.

A huge quantity of papers have been devoted to the study of the distribution of electromagnetic fields in these devices or to their application in different fields of physics and technology (optoelectronic or millimetric devices, metrology, apparatus for spectroscopy). In all these applications the most appealing properties are the manufacture simplicity, the high merit factor Q and the intrinsic wide working spectral range. The counterpart of these abilities is that, when quantitative or precise measurements are requested in different applications, a proper modeling of the electromagnetic field distribution becomes mandatory.

The electromagnetic modeling of WGDRs is typically more difficult in comparison to the modeling of a metallic resonator; indeed, the support of the electromagnetic field in an open dielectric resonator is necessarily unbounded and the irradiation losses always differ from zero **[6]**. As a consequence, the numerical techniques necessary for an accurate modeling should be developed in the complex field using functions defined on unbounded support **[7]**.

Owing to the importance for present and future applications, the focus of this paper is the study of the actual frequency bandwidth of planar cylindrical WG dielectric resonators. Through a simple even if general approach, some limiting properties of a WGDR are first investigated. Starting from the limit radii and modal indices, the working frequency band of a WGDR for a fixed family of resonant modes **[8]** is calculated; the obtained band, called ideal frequency band, only depends on the complex dielectric constants of the involved materials. When however the concurrent presence of other resonant modes is properly taken into account, a reduced operating bandwidth, called effective frequency band, is obtained. The developed analytical procedure gives the connection between the ideal and the effective operating bandwidth; it is indeed demonstrated that the envelope of all the effective frequency bands obtained varying the dimensions of the resonator corresponds to a “parent” frequency band coinciding with the ideal one. Although the extension of the ideal band could span a width of several decades, the extension of the effective frequency band for a given WGDR is at least **k** times narrower (in terms of decades), being **k** the dimension of the space generated by the modal indices. The used procedure and approximations carried out tend to privilege the possibility of grasping the physical meaning and obtaining simple and direct expressions, rather than of deducing a very accurate formulation, being explored by this way all the fundamental aspects. More accurate calculations can be however obtained with a suitable numerical modeling.

The plan of the paper is the following. Sect. 2 presents the general aspects of the propagation of electromagnetic waves in dielectric cylindrical structures, showing some peculiarities of whispering gallery resonators. In particular, the analysis of their irradiation and dielectric losses leads to the calculation of the minimum and maximum allowed values of the azimuthal modal index. Sect. 3 is devoted to the calculation of the ideal frequency band of a WGDR, connected to the limit values calculated in the previous section. Sect. 4 shows that the presence of the transverse modes in the WGDR affects the ideal bandwidth of the resonator leading to a reduced effective bandwidth, first calculated for the case of a disk resonator. The obtained results are then generalized to the case of an annular resonator with arbitrary dimensions. It is then discussed the behavior of extension and position of the effective frequency band inside the ideal one in terms of the geometrical factor. Final remarks and outlooks for applications are discussed in Sect. 5.

The developed analysis gives the necessary background which prelude to the prediction and the experimental verification of multiple-band composite WGDRs, which will be presented in a companion paper.

## 2. A general approach

In its simpler form a planar WGDR is a circular disc made with low loss dielectric isotropic material having a diameter much greater than its thickness, in which the confinement of the radiation is based on the total internal reflection mechanism **[8,9]**. WGDRs support two different families of resonant hybrid modes, usually denoted $WGE_{n,m,l}$ and $WGH_{n,m,l}$; the first ones are quasi-TE modes, with the electric field lying approximately in the plane of the resonator, while the second ones are quasi-TM modes, with the magnetic field lying approximately in the plane of the resonator. The three modal indices **n**, **m** and **l** have the following meanings: **n**, the azimuthal index, is equal to the number of wavelengths in a whole turn around the resonator; **m**, the radial index, is equal to the number of nodes of the energy flux along the radius of the resonator, excluded that one in its center; **l**, the axial index, is equal to the number of nodes of the energy flux along the axis of the resonator. An approximate analytical description of the resonant modes can be found elsewhere **[9,10]**; according to this approach the behavior of the electromagnetic (e.m.) field inside the resonator is given in cylindrical coordinates by an expression of the form:

$$\mathbf{J_n}(\beta\rho)\begin{Bmatrix}\mathbf{\cos(hz)}\\ \mathbf{\sin(hz)}\end{Bmatrix}\mathbf{e}^{\mathbf{i}(\mathbf{n}\varphi+\omega t)} \qquad \mathbf{(1)}$$

where $\mathbf{J_n}(\beta\rho)$ is the Bessel function of the first kind; $\beta$ and **h** are the transverse and axial propagation constants, respectively, related by the equation **[8]**

$$\beta^2 = \mathbf{k}^2 - \mathbf{h}^2 \qquad \mathbf{(2)};$$

here $\mathbf{k} = \frac{\omega}{\mathbf{c}}\sqrt{\varepsilon\mu}$, **c** being the speed of light in vacuum and $\omega$ the angular frequency of the radiation; $\varepsilon$ and $\mu$ are the real parts of the dielectric and magnetic permittivities of the material forming the resonator. In the following the magnetic permittivities will be assumed equal to **1**. All the properties of planar WGDRs investigated will be deduced starting from the above field description; this is equivalent to the assumption that the field distribution of WG modes in planar WGDRs can be obtained by WG modes of a related infinite waveguide, suitably choosing the axial propagation constant as pointed out in **[8, 9]**. As in the case of WG modes of an infinite circular dielectric waveguide **[11]**, special attention will be paid to the case of weak axial propagation constants (**h<<k**): a basic role will then be assigned to the modal index **n**.

The rephasing condition gives for the index **n** the relation **[8]**

$$\mathbf{n} = \beta\rho_{\mathbf{m}} \qquad \mathbf{(3)},$$

where $\rho_{\mathbf{m}}$ is the radius of the so-called modal caustic, defined essentially as the cylindrical surface inside the resonator beyond which the electromagnetic field is confined. In terms of the geometrical optics representation the modal caustic corresponds to the envelope of the rays of the considered resonant mode.

The search for the operating bandwidth of a WGDR must firstly investigate into the realm of *allowed* values of the azimuthal index **n.** According to an application oriented operating approach, a definite value of **n** can be considered allowed when the corresponding field distribution presents losses weak enough to have a resolved spectrum and then a "true" resonant behavior. The main channels of losses for a dielectric resonator are given by the irradiation losses and the dielectric losses of involved materials. As already stated, the confinement of the radiation in a WGDR is based on the total internal reflection mechanism; by this way the irradiation losses, although always different from zero **[6]**, can be reduced to a negligible level with a suitable design of the resonator. To this aim the geometrical dimensions and the regularity of the surfaces play a fundamental role. The total reflection can be obtained on the plane surfaces of the resonator essentially when the optical thickness of the disc is equal or greater than the wavelength in the medium **[8].** The confinement near the curved surface of the disc can be usefully discussed in terms of the proper caustic $r_{\mathbf{p}}$ which corresponds, in terms of the geometrical optics representation, to the envelope of the rays which incide at the limit angle on the curved surface of the resonator. The proper caustic is given by **[12]**

$$r_{\mathbf{p}} = \mathbf{r}\sqrt{\frac{e_{\mathbf{ext}}}{e_{\mathbf{int}}}} \qquad \mathbf{(4)},$$

where $e_{\mathbf{int}}$ and $e_{\mathbf{ext}}$ are the real parts of the dielectric constant of the medium forming and surrounding the resonator, respectively, and **r** is the radius of the resonator.

Since an effective confinement near the curved surface of the disc is obtained when the angle of incidence of the rays on this surface is greater than the limit angle, the caustics must satisfy the following condition

$$\mathbf{r} > r_{\mathbf{m}} > r_{\mathbf{p}} \qquad \mathbf{(5)};$$

in particular the radial irradiation losses decrease when the ratio $r_{\mathbf{m}}/r_{\mathbf{p}}$ increases **[8,13]**. The limit condition for an effective radial field confinement is obtained when $r_{\mathbf{m}}=r_{\mathbf{p}}$. Correspondingly the minimum value for the modal caustic is given by:

$$\left(r_{\mathbf{m}}\right)_{\mathbf{min}} = \mathbf{r}\sqrt{\frac{e_{\mathbf{ext}}}{e_{\mathbf{int}}}} \qquad \mathbf{(6)};$$

as a consequence the minimum value $\mathbf{n}_{\mathbf{min}}$ of the azimuthal index, fixed by the irradiation losses, is equal to (Eq. 3)

$$\mathbf{n}_{\mathbf{min}} = \mathbf{r}\, b\sqrt{\frac{e_{\mathbf{ext}}}{e_{\mathbf{int}}}} \qquad \mathbf{(7)}.$$

On the other hand, a maximum value is imposed to the modal caustic of the resonator and therefore to the azimuthal index by the energy dissipation due to dielectric losses, according to the criterion above discussed. The search of these values can be developed by introducing the absorption coefficient $\alpha$ for the intensity of the e.m. radiation; in the case of free propagation in an infinite homogeneous medium this coefficient is given by **[14]**

$$\alpha = \frac{2\pi}{\lambda}\,\mathbf{tan}\,\delta = \frac{\omega}{\mathbf{v}}\,\mathbf{tan}\,\delta \qquad \mathbf{(8)},$$

where $\lambda$ and $\mathbf{v}$ are the wavelength and the phase velocity of the radiation in the medium, respectively, and $\delta$ is its loss angle. In a bounded medium an effective wavelength $\bar{\lambda}$ as a function of the spatial coordinates can be defined; in the case of a WGDRs the wavelength of interest is related to the circular propagation in the plane of the resonator, where the condition of positive interference must be fulfilled (Eq. 3). Analogously to the propagation in an infinite medium, where $\lambda = \frac{2\pi}{\mathbf{k}}$, the natural definition of a wavelength $\bar{\lambda}$ related to the propagation in the plane of the resonator can be obtained by using the propagation factor $\beta$:

$$\bar{\lambda} = \frac{2\pi}{\beta} \qquad \mathbf{(9)}.$$

The meaning of $\bar{\lambda}$ can be inferred considering the phase factor $\mathbf{e}^{\mathbf{i}\nu\varphi}$ which appears in Eq. 1; combining Eqs. 3 and 9 this factor reduces to $\mathbf{e}^{\mathbf{i}\frac{2\pi}{\bar{\lambda}}\rho_{\mathbf{m}}\varphi}$. This last expression shows that $\bar{\lambda}$ is the distance between two consecutive peaks of the field along the curvilinear coordinate $\rho_{\mathbf{m}}\varphi$, defined on the modal caustic. The wavelength $\bar{\lambda}$ is then the effective wavelength of the radiation measured along the modal caustic of the resonator. Since the planes of constant phase are radial planes (Eq. 1), in order to satisfy the conditions of positive interference for any curvilinear coordinate $\rho\varphi$, the above concept can be generalized by defining an effective wavelength $\lambda_{\mathbf{eff}}(\rho)$ which is everywhere proportional to $\rho$. As a consequence $\lambda_{\mathbf{eff}}(\rho) = \lambda_{\mathbf{eff}}(\rho_{\mathbf{m}})\frac{\rho}{\rho_{\mathbf{m}}} = \bar{\lambda}\frac{\rho}{\rho_{\mathbf{m}}}$. The absorption coefficient $\alpha$ along a circular path with radius $\rho$ and center on the axis of the resonator can be written, analogously to Eq. 8 and using Eqs. 3 and 9, as

$$\alpha(\rho) = \frac{2\pi}{\lambda_{\mathbf{eff}}(\rho)}\,\mathbf{tan}\,\delta = \frac{\nu}{\rho}\,\mathbf{tan}\,\delta \qquad \mathbf{(10).}$$

The damping of the electromagnetic wave can be completely described by Eq. 10, under the approximations of weak axial propagation and of resonant mode completely confined in the resonator, so making negligible the dielectric losses due to the medium external to the resonator.

The normalized wave amplitude after a single turn $\mathbf{A_s}$ along a circular path with radius $\rho$ is given by

$$\mathbf{A_s} = \mathbf{e}^{-2\pi\rho\frac{\alpha(\rho)}{2}} = \mathbf{e}^{-\pi\rho_{\mathbf{m}}\frac{2\pi}{\lambda}\mathbf{tan}\,\delta} \qquad \mathbf{(11),}$$

and results independent of $\rho$. This circumstance allows to develop the following analysis in terms of $\rho_{\mathbf{m}}$ only.

In order to obtain a resolved resonance spectrum it is necessary that the wave resulting from the infinite round trips along the resonator gives an effective interference effect with the input wave. As a definition of resolved spectrum, the Rayleigh criterion is assumed **[15]**. Accordingly, the normalized wave amplitude **A** resulting from all the round trips can assume the minimum value $\mathbf{A_{min}}$ given by $\left(\frac{\mathbf{1 - A_{min}}}{\mathbf{1 + A_{min}}}\right)^{\mathbf{2}} = \mathbf{0.81}$, which corresponds to $\mathbf{A_{min}}$=**0.05263.** The fulfillment of the Rayleigh criterion guarantees that two consecutive resonances of the same family (fixed transverse modal indices $\bar{\mathbf{m}}, \bar{\mathbf{l}}$, consecutive indices **n**) are resolved.

The amplitude **A** is related to the amplitude $\mathbf{A_s}$ by the relation

$$\sum_{j=1}^{\infty} (\mathbf{A_s})^j = \frac{\mathbf{A_s}}{\mathbf{1 - A_s}} = \mathbf{A}\,;$$

since the amplitude $\mathbf{A_{min}}$ corresponds to the maximum allowed value $(\rho_{\mathbf{m}})_{\mathbf{max}}$ of the modal caustic, it follows, by using Eq. 11 and taking into account that $\frac{\mathbf{1}}{\pi}\mathbf{ln}\left(\frac{\mathbf{1 + A_{min}}}{\mathbf{A_{min}}}\right) = \frac{\mathbf{1}}{\pi}\mathbf{ln}(\mathbf{20}) \cong \mathbf{1}$, that

$$(\rho_{\mathbf{m}})_{\mathbf{max}} = \frac{\mathbf{1}}{\pi\beta\,\mathbf{tan}\,\delta}\mathbf{ln}\left(\frac{\mathbf{1 + A_{min}}}{\mathbf{A_{min}}}\right) = \frac{\mathbf{1}}{\beta\,\mathbf{tan}\,\delta} \qquad \mathbf{(12).}$$

The maximum azimuthal index $\mathbf{n_{max}}$, fixed by the dielectric losses, is given by (Eq. 3)

$$\mathbf{n_{max}} = \frac{\mathbf{1}}{\mathbf{tan}\,\delta} \qquad \mathbf{(13).}$$

Under the same approximations leading to Eq. 13, the merit factor **Q** of the resonator is only due to the dielectric losses of the medium forming the resonator, namely $\mathbf{Q} \approx (\mathbf{tan}\,\delta)^{-\mathbf{1}}$; it follows that the procedure used to obtain Eq. 13 is equivalent to assume that the minimum finesse $\Im_{\mathbf{lim}}$ of the family $\overline{\mathbf{m},\boldsymbol{\lambda}}$ satisfies the relation $\Im_{\mathbf{lim}} = \frac{\mathbf{Q}}{\boldsymbol{\nu}_{\mathbf{max}}} = \mathbf{1}$. The self-consistency of the procedure is confirmed since the result reported in Eq. 13, based on the assumption of the Rayleigh criterion, corresponds here to the assumption of unitary finesse which, in turn, is another effective way to define a resolved spectrum.

Eqs. 6, 7 and Eqs. 12, 13 give the limit conditions connected to the irradiation losses (lower bound) and to dielectric losses (upper bound), respectively. The allowed values of $\boldsymbol{\nu}$ belong to the interval $(\boldsymbol{\nu}_{\mathbf{min}}, \boldsymbol{\nu}_{\mathbf{max}})$; the value $\boldsymbol{\nu}_{\mathbf{max}}$ can be reached if in this limit the irradiation losses are negligible in comparison to the dielectric ones, whereas on the contrary the value $\boldsymbol{\nu}_{\mathbf{min}}$ can be reached when in this limit the dielectric losses are negligible in comparison to the irradiation ones**.** A numerical analysis of the behavior of both the irradiation and dielectric losses as a function of $\boldsymbol{\nu}$ can be found in Ref. **[16]**.

## 3. Ideal frequency band of a WGDR

The limits in the operation of a WGDR discussed in Sect. 2 are obtained in a general way; however the dependence of these limits on direct physical quantities like as the resonance frequency $\omega$ and the transverse modal indices **m** and **l** does not appear explicitly and can be inferred only in Eqs 7 and 12, owing to the presence of the propagation constant $\beta$. In the following some approximations will be taken into account in order to achieve directly significant expressions for the obtained results. The ideal frequency bandwidth for a WGDR with given size will then be obtained by considering the fundamental modes only; the effects of transverse modes with modal indexes **m**, **l** different from zero will be discussed in the next section.

First of all, the modal caustic can be conveniently expressed as **[17,18]**

$$\rho_{\mathbf{m}} \approx \begin{cases} \boldsymbol{\rho} - (\mathbf{m}+\mathbf{1})\bar{\lambda} & \mathbf{for} \quad \mathbf{m} \sim \mathbf{1} \\ \boldsymbol{\rho} - (\mathbf{m}+\mathbf{1})\dfrac{\bar{\lambda}}{\mathbf{2}} & \mathbf{for} \quad \mathbf{m} >> \mathbf{1} \end{cases} \qquad \mathbf{(14)}$$

The contribution due to the index **l** (related to the axial propagation) is essentially included in $\bar{\lambda} = \frac{\mathbf{2}\pi}{\beta}$. For low **m** values, the substitution of Eq. 14 into Eq. 6 and Eq. 12 gives, respectively

$$\left(\frac{\mathbf{r}}{\bar{\lambda}}\right)_{\mathbf{min}} = (\mathbf{m}+\mathbf{1})\frac{\sqrt{\varepsilon_{\mathbf{int}}}}{\sqrt{\varepsilon_{\mathbf{int}}}-\sqrt{\varepsilon_{\mathbf{ext}}}} \qquad \textbf{(15)}$$

and

$$\left(\frac{\mathbf{r}}{\bar{\lambda}}\right)_{\mathbf{max}} = \frac{\mathbf{1}}{\mathbf{2}\pi\,\mathbf{tan}\,\delta} + (\mathbf{m}+\mathbf{1}) \qquad \textbf{(16)}.$$

The limit values for **n** can be now directly related to the dielectric properties of the involved materials. For $\mathbf{n_{min}}$, Eqs. 7 and 15 indeed give:

$$\mathbf{n}_{\mathbf{min}} = \mathbf{2}\pi\left(\frac{\mathbf{r}}{\bar{\lambda}}\right)_{\mathbf{min}}\sqrt{\frac{\varepsilon_{\mathbf{ext}}}{\varepsilon_{\mathbf{int}}}} = \mathbf{2}\pi(\mathbf{m}+\mathbf{1})\frac{\sqrt{\varepsilon_{\mathbf{ext}}}}{\sqrt{\varepsilon_{\mathbf{int}}}-\sqrt{\varepsilon_{\mathbf{ext}}}} \qquad \textbf{(7a).}$$

The limit values for the azimuthal index **n**, and then the consistence condition $\mathbf{n_{max}} > \mathbf{n_{min}}$, are now only related to the dielectric properties of the involved materials. In addition, it is worthwhile to note that it is always possible to realize an open dielectric resonator for any positive difference $\sqrt{\varepsilon_{\mathbf{int}}} - \sqrt{\varepsilon_{\mathbf{ext}}}$, provided that $\mathbf{tan}\,\delta$ is low enough.

Eqs. 15 and 16 can be used to calculate, for a fixed $\bar{\lambda}$, the limit values of the radius **r** of a well working WGDR. Viceversa, if the radius **r** is given, the same equations give the maximum value, $\bar{\lambda}_{\mathbf{max}} = \bar{\lambda}(\omega_{\mathbf{min}})$, and the minimum value, $\bar{\lambda}_{\mathbf{min}} = \bar{\lambda}(\omega_{\mathbf{max}})$, being $\omega_{\mathbf{min}}$ and $\omega_{\mathbf{max}}$ the limit values of the resonance frequency. This implies:

$$\frac{\bar{\lambda}(\omega_{\mathbf{min}})}{\bar{\lambda}(\omega_{\mathbf{max}})} = \frac{\beta(\omega_{\mathbf{max}})}{\beta(\omega_{\mathbf{min}})} = \left[\frac{\mathbf{1}}{\mathbf{2}\pi(\mathbf{m}+\mathbf{1})\mathbf{tan}\,\delta} + \mathbf{1}\right]\frac{\sqrt{\varepsilon_{\mathbf{int}}}-\sqrt{\varepsilon_{\mathbf{ext}}}}{\sqrt{\varepsilon_{\mathbf{int}}}} \qquad \textbf{(17).}$$

If the case of weak axial propagation **h<<k** is considered, $\beta \rightarrow \mathbf{k}$ and $\bar{\lambda}(\omega)$ tends to the usual wavelength $\lambda(\omega) = \dfrac{\mathbf{2}\pi\mathbf{c}}{\omega\sqrt{\varepsilon_{\mathbf{int}}}}$. For the fundamental modes $WGM_{n,0,0}$ Eq. 17 then reduces to

$$\frac{\omega_{\mathbf{max}}}{\omega_{\mathbf{min}}} = \left(\frac{\mathbf{1}}{\mathbf{2}\pi\,\mathbf{tan}\,\delta} + \mathbf{1}\right)\frac{\sqrt{\varepsilon_{\mathbf{int}}}-\sqrt{\varepsilon_{\mathbf{ext}}}}{\sqrt{\varepsilon_{\mathbf{int}}}} \qquad \textbf{(18),}$$

that holds if the involved dielectric constants are independent of the frequency in the interval $(\omega_{\mathbf{min}}, \omega_{\mathbf{max}})$. If the dependence on the frequency of the complex dielectric constants cannot be neglected, the limit frequencies can be obtained by an iterative procedure from the relations $\omega_{\mathbf{max}} = \mathbf{f}(\varepsilon(\omega))$ and $\omega_{\mathbf{min}} = \mathbf{g}(\varepsilon(\omega))$, where **f** and **g** express the functional dependence of $\omega_{\mathbf{max}}$ and $\omega_{\mathbf{min}}$ from the dielectric constants, respectively. This iterative procedure must continue until the convergence is reached or until the condition of consistence $\mathbf{n_{max}}(\omega) \geq \mathbf{n_{min}}(\omega)$ fails. Eq. 18 gives the width of the frequency band in terms of the ratio between the limit frequencies; such ratio will be called in the following frequency band extension.

The above analysis gives the ratio between the limit frequencies only in terms of the complex dielectric permittivities of the involved materials. In particular a considerable role is plaid by the intrinsic losses of the material forming the WGDR. As a practical example, assuming $\frac{\varepsilon_{\mathbf{int}}}{\varepsilon_{\mathbf{ext}}} \geq \mathbf{2}$, from Eq. 18 it follows that the ideal band becomes wider than six decades when materials with $(\tan\delta) \leq 10^{-8}$ are employed (like sapphire or rutile at microwave frequencies cooled down to the liquid helium temperature).

## 4. Effective frequency band of a WGDR

The ideal frequency band cannot be obtained in elementary structures like single discs or spherical dielectric resonators; the fundamental reason that limits the real frequency band (in the following called effective frequency band) is related to the existence of transverse modes with modal indices **m** and/or **l** different from zero. In order to clarify the influence of these modes on the ideal frequency band, the illustrative case of a simple disc WGDR is first discussed. Afterwards the more general case of a circular ring WGDR with arbitrary internal radius and thickness will be treated.

### *4a. Simple disc WGDR*

The number **L** of modes with different axial index **l** supported by a dielectric disc is essentially fixed by its thickness **t.** In order to calculate **L**, $\beta$ can be rewritten, by using Eq. 2, as

$$\beta = \sqrt{\mathbf{k}^2 - \mathbf{h}^2} = \sqrt{\varepsilon\frac{\omega^2}{\mathbf{c}^2} - \left[\frac{\pi(\mathbf{l}+\mathbf{1})}{\mathbf{t}}\gamma\right]^2} \approx \sqrt{\varepsilon\frac{\omega^2}{\mathbf{c}^2} - \left[\frac{\pi(\mathbf{l}+\mathbf{1})}{\mathbf{t}}\right]^2} \qquad \mathbf{(19)},$$

where the axial propagation constant **h** has been expressed in terms of index **l** and thickness **t** of the resonator; the factor $\gamma$, typically of the order of unity, becomes equal to 1 when the field goes to zero on the plane surfaces of the resonator, as in the case of a metallic cavity **[19]**.

The number **L** can be obtained from Eq. 19, by imposing an angle of incidence of the radiation on the plane surfaces of the resonator greater than the critical one **[8]**. The values of the axial propagation constant are then limited by $\mathbf{h}_{\mathbf{max}}$, which satisfies the equalities $\left(\mathbf{1}-\frac{\varepsilon_{ext}}{\varepsilon_{int}}\right)\cdot \mathbf{k}^2 = \mathbf{h}^2_{\mathbf{max}} = \left(\frac{\pi \mathbf{L}}{\boldsymbol{\tau}}\right)^2$ **[8]**; **L** is then given by $\mathbf{L} = \mathbf{2}\cdot\sqrt{\mathbf{1}-\frac{\varepsilon_{ext}}{\varepsilon_{int}}}\cdot\frac{\boldsymbol{\tau}}{\lambda} = \chi\frac{\boldsymbol{\tau}}{\lambda}$. The number **M** of modes with different radial index **m** is related to the radius **r** and can be estimated from Eq. 14 as $\mathbf{M} \approx \frac{(\mathbf{r}-\rho_{\mathbf{p}})}{\lambda}$, under the condition given in Eq. 5. In the spirit of the present approach, the corrections to **M** due to the limit $\mathbf{r}-\rho_{\mathbf{p}} >> \lambda$ in Eq. 14, as well as the difference between $\lambda$ and $\bar{\lambda}$, are neglected: these two approximations tend however to compensate each other. The conditions **L<1** and/or **M<1** correspond to a structure with high irradiation losses for radiation with wavelength $\lambda$.

In addition to the fundamental modes $WGM_{n,0,0}$, characterized by a free spectral range $\Delta\omega_{\mathbf{fsr}}$, all transverse modes $WGM_{n,m,l}$, obtained varying the index **n** for each allowed radial and axial index, must be considered. Assuming the free spectral range of these transverse modes similar to that of the fundamental ones, the total number $\mathbf{N}_{\mathbf{fsr}}$ of resonances in a given interval $\Delta\omega_{\mathbf{fsr}}$ around $\omega$ is given by

$$\mathbf{N}_{\mathbf{fsr}}(\omega) \approx \mathbf{L}\cdot\mathbf{M} \approx \mathbf{2}\boldsymbol{\tau}\cdot\mathbf{r}\frac{\sqrt{\varepsilon_{int}-\varepsilon_{ext}}}{\varepsilon_{int}}\frac{\sqrt{\varepsilon_{int}}-\sqrt{\varepsilon_{ext}}}{\lambda^2} \propto \omega^2 \qquad \mathbf{(20).}$$

When the frequency increases, the spectrum of the resonator becomes more and more dense until the structure is no more resonant, at least if it is not possible to excite selectively the different modes; from the point of view of the geometrical optics, when the dimensions of the resonator become too large in respect to the wavelength, the number of the possible paths by which a ray returns to its starting point exceedingly increases, and it is no more possible to find an univocal condition for constructive interference. The presence of the transverse modes then limits the effective frequency band of the resonator in respect to the ideal one; this limitation can be expressed in terms of the effective *finesse* of the resonator, given by $\Im_{\mathbf{eff}} = \frac{\Delta\omega_{\mathbf{fsr}}}{\sum'_{\mathbf{n,m,l}}\Delta\omega^{\mathbf{(n,m,l)}}_{\mathbf{res}}} = \frac{\Delta\omega_{\mathbf{fsr}}}{\mathbf{N}_{\mathbf{fsr}}\cdot\Delta\omega_{\mathbf{res}}}$, where the symbol $\sum_{\nu,m,l}^{l}$ indicates the sum over all resonances lying in an interval $\Delta\omega_{fsr}$ wide, $\Delta\omega_{res}^{(\nu,m,l)}$ is the full width at half maximum (FWHM) of the resonances labeled by **(n,m,l)**, and $\Delta\omega_{res}$ is the resulting mean FWHM. As already discussed, the spectrum of the resonator can be considered resolved if $\Im_{eff} \geq 1$. For WGDRs designed with maximum bandwidth starting from $\omega_{min}$, the dimensions must be chosen in order that near $\omega_{min}$ the spectrum has only modes of the form $WGM_{n,0,0}$, with $\nu \sim \nu_{min}$; this implies **L** = **M** = 1, so that from Eq. 20 it follows $N_{fsr} = 1$. As a consequence

$$N_{fsr}(\omega) = \left(\frac{\omega}{\omega_{min}}\right)^2 \qquad \textbf{(21).}$$

Let us assume that $\frac{\Delta\omega_{fsr}}{\Delta\omega_{res}}$ coincides with the *finesse* of the fundamental modes $\Im = \frac{Q}{\nu} \approx \frac{(\tan\delta)^{-1}}{\nu}$; the upper limit $\omega_{upp}$ of the effective frequency band, limited by the overlap between different modes, can be obtained by the condition $\Im_{eff}(\omega_{upp}) = \frac{(\tan\delta)^{-1}}{\nu(\omega_{upp})}\frac{1}{N_{fsr}(\omega_{upp})} = 1$, which gives

$$\left(\frac{\omega_{upp}}{\omega_{min}}\right)^2 = \frac{(\tan\delta)^{-1}}{\nu(\omega_{upp})} \qquad \textbf{(22).}$$

Neglecting the axial propagation constant **h**, the azimuthal index $\nu$ for the fundamental modes can be expressed, by using Eqs. 3 and 14, as $\nu(\omega) = \frac{\omega\sqrt{\varepsilon_{int}}}{c}[\rho - \lambda(\omega)]$. Eq. 3, for a fixed radius $\rho$, and Eq. 6, give the minimum azimuthal index $\nu_{min}(\omega_{min}) = \frac{\omega_{min}\sqrt{\varepsilon_{ext}}}{c}\rho$; the ratio $\frac{\nu(\omega)}{\nu(\omega_{min})}$ then gives

$$\nu(\omega) = \nu_{min}\frac{\omega}{\omega_{min}}\sqrt{\frac{\varepsilon_{int}}{\varepsilon_{ext}}}\left[1 - \frac{\lambda(\omega)}{\rho}\right] \qquad \textbf{(23).}$$

The ratio $\frac{\lambda(\omega)}{\rho}$ can be expressed in terms of $\frac{\omega}{\omega_{min}}$ by substituting the value of $\rho$ with its expression derived from Eq. 15, which gives $\frac{\lambda(\omega)}{\rho} = \frac{\omega_{min}}{\omega}\frac{\sqrt{\varepsilon_{int}} - \sqrt{\varepsilon_{ext}}}{\sqrt{\varepsilon_{int}}}$; making use of this expression and of Eq. 7b and Eq. 23, Eq. 22 gives:

$$\left(\frac{\omega_{upp}}{\omega_{min}}\right)^3 \left(1 - \frac{\omega_{min}}{\omega_{upp}}\frac{\sqrt{\varepsilon_{int}} - \sqrt{\varepsilon_{ext}}}{\sqrt{\varepsilon_{int}}}\right) =$$

$$= \frac{(\tan\delta)^{-1}}{2\pi}\left(\frac{\sqrt{\varepsilon_{int}} - \sqrt{\varepsilon_{ext}}}{\sqrt{\varepsilon_{int}}}\right) \approx \frac{\omega_{max}}{\omega_{min}} \qquad \textbf{(24).}$$

When the dominant term in the left hand of Eq. 24 is that of third order, an equivalent form of this equation can be inferred directly from Eq. 19; indeed, since the number of modes increases with the number of the allowed terns $(\mathbf{n},\mathbf{m},\mathbf{l})$, the condition of overlapping spectrum is reached as fast as $\omega^{\mathbf{k}}$, where **k** is the dimension of the space generated by the modal indices.

### *4b. Circular ring WGDR*

The analysis developed in subsection 4a can be generalized to the case of a disc with a central axial hole; this topology is important because it allows a partial suppression of transverse radial modes ($\mathbf{m} \neq 0$).

Let us consider a circular ring WGDR with axial thickness **t** and radial thickness $\Delta\mathbf{r}$, as indicated in Fig. 1.

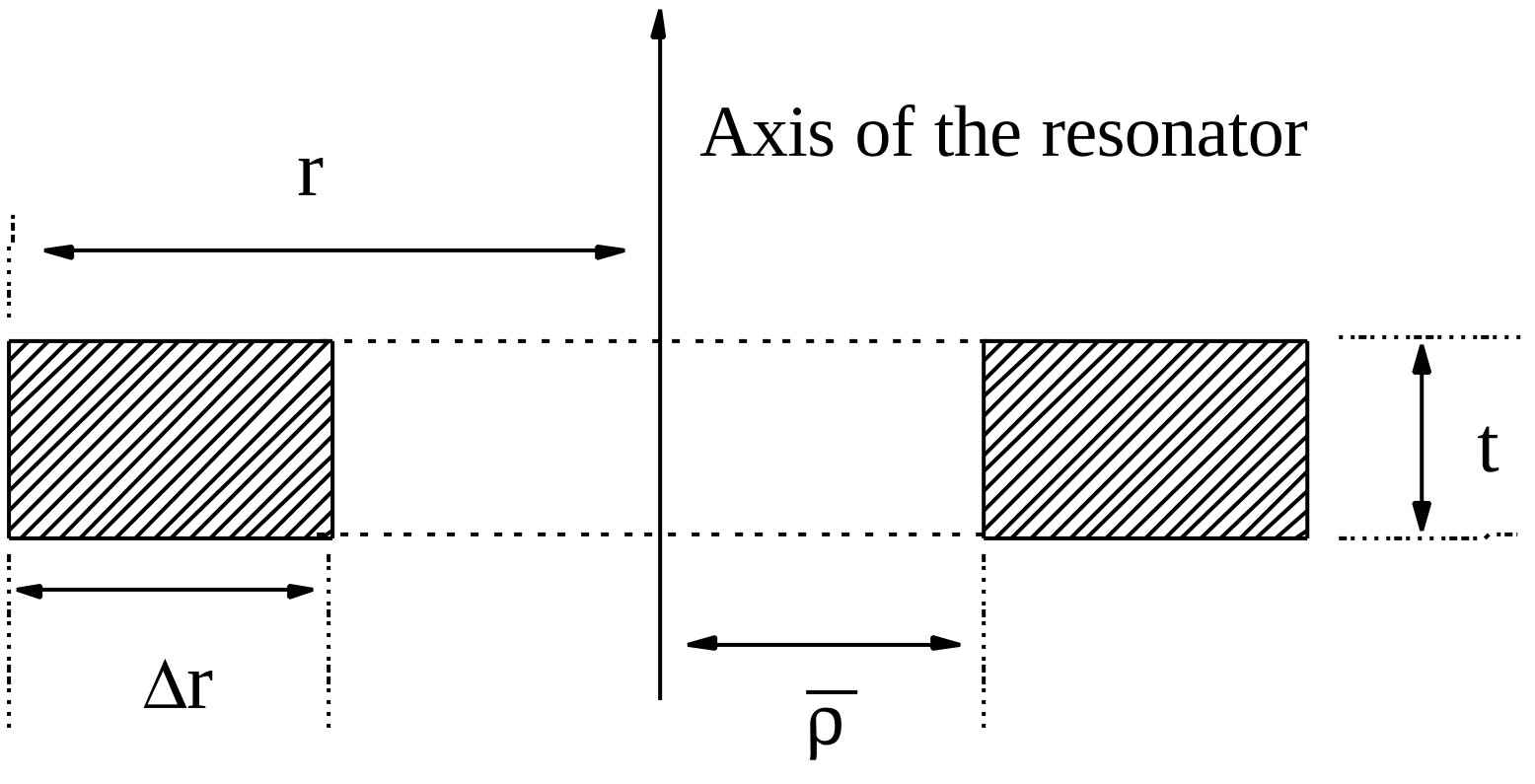

***Fig. 1*** *Schematic drawing of a circular ring dielectric resonator.*

These thicknesses can be expressed in terms of the wavelength $\lambda(\omega_{\mathbf{min}})$ by introducing the parameters $\mathbf{p}$ and $\mathbf{q}$, defined by $\mathbf{t} = \mathbf{p} \cdot \frac{\lambda(\omega_{\mathbf{min}})}{\chi}$ and $\Delta\mathbf{r} = \mathbf{r} - \bar{\rho} = \mathbf{q} \cdot \lambda(\omega_{\mathbf{min}})$, being $\bar{\rho}$ the radius of the central hole; the term $\chi \cdot \mathbf{t}$ will be defined as effective axial thickness. As previously discussed, the irradiation losses become very high for radiation with wavelength greater than $\chi \cdot \mathbf{t}$ and/or $\Delta\mathbf{r}$; all the corresponding resonant modes are then practically forbidden. The lower allowed frequency $\omega_{\mathbf{low}}$ for the resonator of Fig. 1 is the minimum frequency which allows at least one wavelength in both the effective axial and radial thickness; it follows then $\lambda(\omega_{\mathbf{low}}) = \mathbf{min}[\chi \cdot \mathbf{t}, \Delta\mathbf{r}] = \lambda(\omega_{\mathbf{min}})\mathbf{min}[\mathbf{p},\mathbf{q}]$, so that

$$\omega_{\mathbf{low}} = \frac{\omega_{\mathbf{min}}}{\mathbf{min}[\mathbf{p},\mathbf{q}]} \qquad \mathbf{(25);}$$

on the other hand it must be $\omega_{\mathbf{low}} \le \omega_{\mathbf{max}}$, and from Eq. 25 it follows that $\mathbf{min}[\mathbf{p},\mathbf{q}] \ge \frac{\omega_{\mathbf{min}}}{\omega_{\mathbf{max}}}$, which implies $\mathbf{p} \ge \frac{\omega_{\mathbf{min}}}{\omega_{\mathbf{max}}}$ and $\mathbf{q} \ge \frac{\omega_{\mathbf{min}}}{\omega_{\mathbf{max}}}$. The presence of a central hole affects the distribution of the electromagnetic fields only if $\bar{\rho}$ is greater than the proper caustic $\rho_{\mathbf{p}}$; the condition $\bar{\rho} = \rho_{\mathbf{p}}$ then gives the maximum physically meaningful value $\mathbf{q_{max}}$ for $\mathbf{q}$; since $\lambda(\omega_{\mathbf{min}}) = \mathbf{r} - \rho_{\mathbf{p}}$, it results $\mathbf{q_{max} = 1}$. In turn, as $\mathbf{min}[\mathbf{p},\mathbf{q}] \le \mathbf{q_{max}} = \mathbf{1}$, this implies that $\omega_{\mathbf{low}} \ge \omega_{\mathbf{min}}$. The maximum value of $\mathbf{p}$ will be obtained in the following by imposing an unitary effective frequency band.

Being the minimum value between $\chi \cdot \mathbf{t}$ and $\Delta\mathbf{r}$ equal to $\lambda(\omega_{\mathbf{low}})$, its maximum value can be expressed by $\mathbf{max}[\chi \cdot \mathbf{t}, \Delta\mathbf{r}] = \frac{\mathbf{max}[\chi \cdot \mathbf{t}, \Delta\mathbf{r}]}{\mathbf{min}[\chi \cdot \mathbf{t}, \Delta\mathbf{r}]}\lambda(\omega_{\mathbf{low}}) = \frac{\mathbf{max}[\mathbf{p},\mathbf{q}]}{\mathbf{min}[\mathbf{p},\mathbf{q}]}\lambda(\omega_{\mathbf{low}})$; from the definitions of $\mathbf{N_{fsr}}(\omega)$, $\mathbf{L}$ and $\mathbf{M}$, it results that the factor $\frac{\mathbf{max}[\mathbf{p},\mathbf{q}]}{\mathbf{min}[\mathbf{p},\mathbf{q}]}$ represents the total number of allowed modes $\mathbf{N_{fsr}}(\omega_{\mathbf{low}})$.

Since $\mathbf{N_{fsr}}(\omega)$ increases as $\omega^2$, it follows:

$$\mathbf{N_{fsr}}(\omega) = \left(\frac{\omega}{\omega_{\mathbf{low}}}\right)^2 \frac{\mathbf{max}[\mathbf{p},\mathbf{q}]}{\mathbf{min}[\mathbf{p},\mathbf{q}]} \qquad \mathbf{(26).}$$

The condition $\Im_{\mathbf{eff}}(\omega_{\mathbf{upp}}) = \mathbf{1}$ leads now to

$$\left(\frac{\omega_{upp}}{\omega_{low}}\right)^2 \frac{\max[p,q]}{\min[p,q]} = \frac{(\tan\delta)^{-1}}{n(\omega_{upp})};$$

by expressing $\omega_{min}$ in terms of $\omega_{low}$ in Eq. 23, Eq. 24 can be generalized as

$$\left(\frac{\omega_{upp}}{\omega_{low}}\right)^3 \left(1-\min[p,q]\frac{\omega_{low}}{\omega_{upp}}\frac{\sqrt{\varepsilon_{int}}-\sqrt{\varepsilon_{ext}}}{\sqrt{\varepsilon_{int}}}\right) = \frac{(\tan\delta)^{-1}}{2\pi}\left(\frac{\sqrt{\varepsilon_{int}}-\sqrt{\varepsilon_{ext}}}{\sqrt{\varepsilon_{int}}}\right)\frac{(\min[p,q])^2}{\max[p,q]} \qquad \textbf{(27)},$$

Since in Eq. 27 the quadratic term in $\frac{\omega_{upp}}{\omega_{low}}$ is negligible (excluding the marginal condition $\frac{\omega_{upp}}{\omega_{min}} \cong 1$), it is possible to relate the extension of the effective frequency band with that of the ideal one; by using the last equality in Eq. 24, Eq. 27 becomes

$$\frac{\omega_{upp}}{\omega_{low}} \approx \left(\frac{\omega_{max}}{\omega_{min}}\right)^{\frac{1}{3}} \left(\frac{(\min[p,q])^2}{\max[p,q]}\right)^{\frac{1}{3}} \qquad \textbf{(28).}$$

Under these approximations the effective band extension $\frac{\omega_{upp}}{\omega_{low}}$ factorizes in a term given by the dielectric properties of the involved materials (the ideal band extension) and a term given by the geometrical properties. The properties of the effective band can be then obtained, being fixed the dielectric properties of the involved materials, from the geometrical factor $\frac{(\min[p,q])^2}{\max[p,q]}$. In order to maximize this quantity, and then the extension of the effective band, the geometrical factor can be rewritten as the product $\frac{\min[p,q]}{\max[p,q]} \times \min[p,q]$; the first factor is $\leq 1$, and assumes the maximum value for all the pairs $(p,q)$ with $p=q$; also the second factor is $\leq 1$ and assumes its maximum value for all the pairs ($p \geq 1$, $q=1$). Finally, only the pair ($p=1$, $q=1$) simultaneously gives the maximum of both factors and then the maximum of the investigated function. As a consequence, the effective band extension $\frac{\omega_{upp}}{\omega_{low}}$ is limited in respect to the ideal one $\frac{\omega_{\mathbf{max}}}{\omega_{\mathbf{min}}}$ by a factor 3 or greater in terms of decades. Furthermore, the condition $\mathbf{p} = \mathbf{q} = \mathbf{1}$ leads to Eq. 24, as previously stated.

The explicit value of the upper frequency $\omega_{\mathbf{upp}}$ can be obtained combining Eq. 28 and Eq. 25; it follows $\frac{\omega_{\mathbf{upp}}}{\omega_{\mathbf{min}}} = \left(\frac{\omega_{\mathbf{max}}}{\omega_{\mathbf{min}}}\right)^{\frac{1}{3}} \left(\frac{\mathbf{1}}{\mathbf{min[\,p,q]} \cdot \mathbf{max[p,q]}}\right)^{\frac{1}{3}} = \left(\frac{\omega_{\mathbf{max}}}{\omega_{\mathbf{min}}}\right)^{\frac{1}{3}} \left(\frac{\mathbf{1}}{\mathbf{p} \cdot \mathbf{q}}\right)^{\frac{1}{3}}$, so that, remembering that $\mathbf{p} \geq \frac{\omega_{\mathbf{min}}}{\omega_{\mathbf{max}}}$ and $\mathbf{q} \geq \frac{\omega_{\mathbf{min}}}{\omega_{\mathbf{max}}}$, $\omega_{\mathbf{upp}} \leq \omega_{\mathbf{max}}$ and the upper limit of the ideal band $\omega_{\mathbf{upp}} = \omega_{\mathbf{max}}$ is reached if and only if $\mathbf{p} = \frac{\omega_{\mathbf{min}}}{\omega_{\mathbf{max}}}$ and $\mathbf{q} = \frac{\omega_{\mathbf{min}}}{\omega_{\mathbf{max}}}$. The two limit frequencies $\omega_{\mathbf{upp}}$ and $\omega_{\mathbf{low}}$ can then be varied in the whole ideal interval; for a fixed $\omega_{\mathbf{low}}$ (i.e. for a fixed **min[p,q],** see Eq. 25), the maximum effective band extension, related now to the ratio $\frac{\mathbf{min[\,p,q]}}{\mathbf{max[p,q]}}$, is obtained when **p=q**. The effective band extension degenerates however to **1** when the upper frequency $\omega_{\mathbf{upp}}$ reaches the ideal limit $\omega_{\mathbf{max}}$, since in this limit also $\omega_{\mathbf{low}}$ reach $\omega_{\mathbf{max}}$ (Eq. 25).

The previous discussion evidences that the ideal band results from the envelope of the effective bands obtained for all the allowed geometrical dimensions of an annular WGDR.

Finally, by imposing the condition that the effective band is not degenerate, the allowed values for the parameters **p** and **q** are obtained; this condition can be expressed (Eq. 28) as $\frac{\omega_{\mathbf{max}}}{\omega_{\mathbf{min}}} \geq \frac{\mathbf{max[p,q]}}{(\mathbf{min[\,p,q]})^2}$; for **p<q** this leads to the inequality $\mathbf{p}^2 \geq \mathbf{q}\frac{\omega_{\mathbf{min}}}{\omega_{\mathbf{max}}}$, whereas for $\mathbf{p} \geq \mathbf{q}$ the inequality $\mathbf{p} \leq \mathbf{q}^2 \frac{\omega_{\mathbf{max}}}{\omega_{\mathbf{min}}}$ is obtained; the latter inequality also gives the maximum allowed **p** value, which is $\mathbf{p}_{\mathbf{max}} = \frac{\omega_{\mathbf{max}}}{\omega_{\mathbf{min}}}$.

## 5. Concluding remarks and outlooks

In conclusion, the role that geometrical and dielectric properties of a planar WGDR plays in the determination of some basic parameters has been investigated in this paper. In particular, the maximum and minimum allowed radius and azimuthal index have been obtained in terms of complex dielectric constants and working frequency. These results give simple and direct criteria for the design of WGDRs for the different applications, in presence of materials having whatever value of complex dielectric constant.

When the radius of the resonator is fixed, the above discussed analysis gives the limit values of the resonance frequency, so that the working frequency band can be determined in terms of the dielectric properties of the involved materials only. Since the WGDRs are typically characterized by high values of the azimuthal index **n**, they can be considered wideband resonators as the working frequency band is expected much greater than the free spectral range. The obtained results clarify quantitatively the meaning of this general property; the ideal band of extremely low loss materials can span several decades, as for instance in the case of sapphire. However different practical reasons, as the difficulty to find a dielectric material transparent over the full ideal band, severely limit the realization of a very wideband WGDR. Moreover, since in a homogeneous structure the behavior of the different resonant modes is expected quite similar, the selective excitation of only one family of modes is difficult, so the effect of all the active resonant modes must be taken into account. This leads to the meaningful definition of an effective frequency band, which results smaller than the ideal one at least by a factor 3 in terms of decades.

To overcome the limitations due to the presence of transverse modes, non-homogeneous structures can be considered. Some non-homogeneous resonators will be discussed in a following companion paper, in which the multiple-band properties of stacked WGDRs working at their ideal limiting frequencies are substantiated.

**Acknowledgment.** This investigation was partially supported by NATO Grant PST.CLG.976444.